\documentclass{article}

\usepackage{arxiv}

\usepackage[utf8]{inputenc} 
\usepackage[T1]{fontenc}    
\usepackage{hyperref}       
\usepackage{url}            
\usepackage{booktabs}       
\usepackage{amsfonts}       
\usepackage{nicefrac}       
\usepackage{microtype}      
\usepackage{lipsum}		
\usepackage{graphicx}
\usepackage[bf]{caption}

\usepackage{doi}
\usepackage{filecontents}
\usepackage{etoolbox}
\usepackage[natbibapa]{apacite}
\AtBeginDocument{}
\renewcommand{\APACrefnote}[1]{}

\newtoggle{bibdoi}
\newtoggle{biburl}
\makeatletter
\newsavebox{\bib@url}
\newsavebox{\bib@doi}

\undef{\APACrefURL}
\undef{\endAPACrefURL}
\undef{\APACrefDOI}
\undef{\endAPACrefDOI}

\newenvironment{APACrefURL}
  {\global\toggletrue{biburl}\lrbox\bib@url}
  {\endlrbox}

\newenvironment{APACrefDOI}
  {\global\toggletrue{bibdoi}\lrbox\bib@doi}
  {\endlrbox}

\newcommand{\printinfo}{
  \iftoggle{bibdoi}{\usebox{\bib@doi}}{\usebox{\bib@url}}
  \togglefalse{bibdoi}
}

\AtBeginEnvironment{thebibliography}{
\pretocmd{\PrintBackRefs}{%
  \iftoggle{bibdoi}
    {\iftoggle{biburl}{\unskip\unskip}{}\usebox{\bib@doi}}
    {\iftoggle{biburl}{Retrieved from \usebox{\bib@url}}}{}
  \togglefalse{bibdoi}\togglefalse{biburl}%
}{}{}}
\makeatother

\title{Specifying Evacuation Return and Home-switch Stability During Short-term Disaster Recovery\\ Using Location-based Data}

\date{} 					

\renewcommand{\headeright}{}
\renewcommand{\undertitle}{}
\renewcommand{\shorttitle}{}

\hypersetup{
pdftitle={Specifying Evacuation Return and Home-switch Stability During Short-term Disaster Recovery Using Location-based Data},
pdfsubject={},
}

\begin{document}
\maketitle

\begin{center}
{\Large
Cheng-Chun Lee\textsuperscript{a,*},
Charles Chou\textsuperscript{b},
Ali Mostafavi\textsuperscript{a}
\par}

\bigskip
\textsuperscript{a}Urban Resilience.AI Lab, Zachry Department of Civil and Environmental Engineering,\\ Texas A\&M University, 199 Spence St., College Station, TX 77843\\
\vspace{6pt}
\textsuperscript{b} Department of Computer Science and Engineering,\\ Texas A\&M University, 435 Nagle St., College Station, TX 77843\\
\vspace{6pt}
\textsuperscript{*} correseponding author, email: ccbarrylee@tamu.edu
\\
\end{center}
\bigskip
\begin{abstract}
The objectives of this study are: (1) to specify evacuation return and home-switch stability as two critical milestones of short-term recovery during and in the aftermath of disasters; and (2) to understand the disparities among subpopulations in the duration of these critical recovery milestones. Using privacy-preserving fine-resolution location-based data, we examine evacuation return and home move-out rates in Harris County, Texas in the context of the 2017 Hurricane Harvey. For each of the two critical recovery milestones, the results reveal the areas with short- and long-return durations and enable evaluating disparities in evacuation return and home-switch stability patterns. In fact, a shorter duration of critical recovery milestone indicators in flooded areas is not necessarily a positive indication. Shorter evacuation return could be due to barriers to evacuation and shorter home move-out rate return for lower-income residents is associated with living in rental homes. In addition, skewed and non-uniform recovery patterns for both the evacuation return and home-switch stability were observed in all subpopulation groups. All return patterns show a two-phase return progress pattern. The findings could inform disaster managers and public officials to perform recovery monitoring and resource allocation in a more proactive, data-driven, and equitable manner.
\end{abstract}


\section{Introduction}
The objectives of this study are: (1) to specify evacuation return and home-switch stability as two critical milestones of short-term recovery during and in the aftermath of disasters; and (2) to understand the presence of disparities among subpopulations in duration of these critical recovery milestones. The intensity and frequency of extreme weather events—flooding, winter storms, and hurricanes—have increased in the past few decades \citep{coronese_evidence_2019, kryvasheyeu_rapid_2016, stocker_climate_2014}. Monitoring the recovery from these extreme weather events enables determination of whether people have returned to their pre-disaster life and prepared for the following event. The literature \citep{johnson_synthesis_2012, kapucu_collaborative_2014, platt_factors_2018, rouhanizadeh_identification_2019} has recognized that community disaster recovery is a complex process that involves several factors, including resource allocation, population vulnerability, and infrastructure resilience. Yet the assessments of recovery stages are often descriptive, subjective, and lack quantitative and data-driven measures to assess and proactively monitor the progress of community disaster recovery to inform recovery implementation and resource allocation \citep{horney_developing_2017, olshansky_evolution_2014, sledge_disaster_2019}. Hence, departing from the standard approach for designation of recovery stages descriptively, we aim to determine critical milestones based on population activity patterns embedded in location-based big data, which is crucial to help decision makers and responders understand and monitor community recovery progress and allocate resources to communities. \\

In this study, critical recovery milestones are defined as times at which community functionality, such as life activities and commerce, return to steady state. This study focuses on two critical milestones of short-term recovery during and in the aftermath of disasters: evacuation return and home-switch stability. During disasters, people may evacuate to escape life-threatening circumstances. Persons severely impacted by disasters who fail to evacuate may incur physical health issues as well as long-term mental health problems \citep{few_flood_2013, lane_health_2013, suzuki_impact_2021}. Houses damaged by the disaster may need repair before they are again inhabitable, forcing residents to relocate until their homes are inhabitable. Thus, by employing evacuation return and home-switch stability as indicators of short-term community disaster recovery, we can enable the understanding of community short-term recovery progress at fine scales. Also, specifying and monitoring these short-term critical recovery milestones reveals the trajectory of long-term recovery for some subpopulations. The evacuation return and home-switch stability can help decision-makers monitor community disaster recovery progress proactively, reveal disparate recovery progress in subpopulations, understand population responses to disasters, and reduce the influence on health and the local economy for future extreme weather events.\\

With location-based data, several studies have examined population mobility during disasters \citep{coleman_human_2021, gray_natural_2012, hsu_limitations_2021, lu_predictability_2012, pastor-escuredo_flooding_2014, yabe_mobile_2019} and assessed disaster impacts \citep{bonaccorsi_economic_2020, esmalian_characterizing_2022, fan_network_2020, lee_community-scale_2021, wang_quantifying_2014, yuan_unraveling_2022}; however, the majority of these studies focus on evacuation patterns \citep{deng_high-resolution_2021, song_prediction_2016}, disruption in mobility \citep{arrighi_preparedness_2019, esmalian_disruption_2021, galeazzi_human_2021}, and mobility resilience \citep{fan_evaluating_2021, fan_neural_2021, roy_quantifying_2019, wang_aggregated_2017}. Despite the recognition that population mobility and disaster impact extent are important factors in community resilience and recovery in disasters, few studies have attempted to characterize the disaster recovery process based on patterns of population activities. Currently, data for assessing disaster recovery is collected via public surveys from respondents who experienced disaster events \citep{mitsova_effects_2019}. Compiling and analyzing survey data has significant lag and puts the burden of providing information on affected people. Location-based big data, on the other hand, provides opportunities to investigate the post-disaster recovery on a much finer scale and in a timely manner. Recognizing this, a number of recent studies have examined disaster recovery using location-based data. \citet{yabe_understanding_2020} observed human mobility for five extreme events with more than 1.9 million mobile phone users to examine macroscopic population mobility recovery patterns with consideration of connectedness to neighboring cities and house damage levels. Despite these efforts, little attention has been paid to examining short-term critical recovery milestones in terms of evacuation return and home-switch stability by using location-based big data.\\

\begin{figure}
	\centering
    \includegraphics[width=0.85\linewidth]{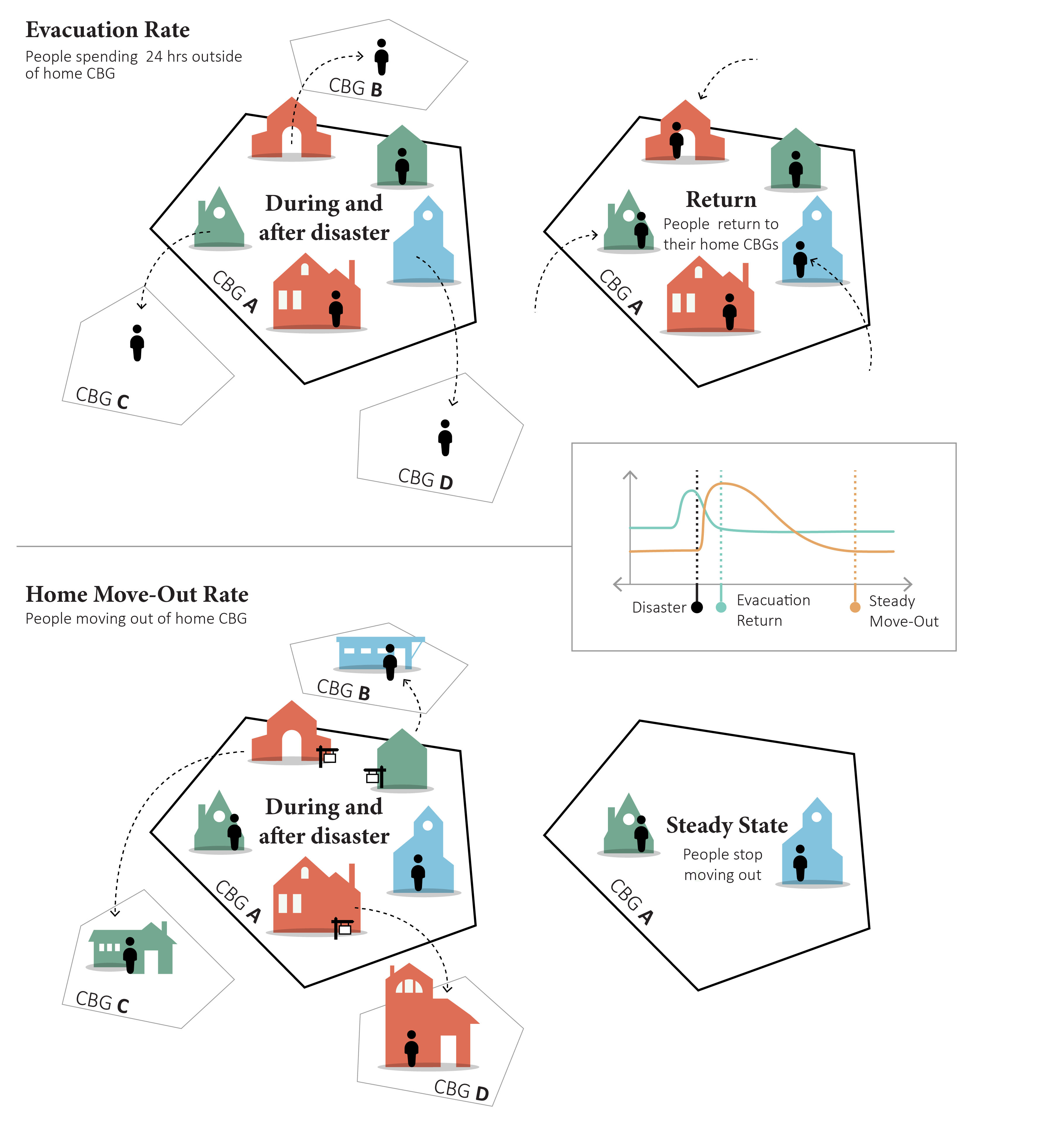}
    \caption{Schematic illustration of two critical community recovery milestones, evacuation return and home-switch stability assessing by evacuation rate and home move-out.}
	\label{fig:fig1}
\end{figure}

In this study, we utilized aggregated location-based data to assess critical community recovery milestones, evacuation return, and home-switch stability, in the aftermath of the 2017 Hurricane Harvey in Harris County, Texas. As shown in Figure \ref{fig:fig1}, two important indices, evacuation and home move-out rates, are defined and examined as indicators for the short-term community recovery milestones of evacuation return and home-switch stability. The return of evacuation rate indicates that people have returned to their homes after evacuation. This milestone indicates that the hazard impacts (such as road inundations and power outages) have diminished, and residents felt safe to return to their homes. On the other hand, a steady state of home move-out rate represents that people ceased moving out of their home census block group (CBG), and the community has returned to stasis in terms of home switch. A greater than normal home CBG switch could suggest that residents whose homes were impacted are moving to other areas. The return of home-switch rate to a stable state suggest that impacted residents have found a new residence (new permanent home or temporary home). There are various reasons for switching homes after disasters; for example, in the context of flooding events, people may decide to sell their homes due to the lack of flood insurance, or people may want to relocate to places of relatively higher elevations to avoid future flooding impacts. Also, people who want to repair their current homes may need to move to temporary quarters while their damaged homes are under repair and restoration. People living in rental homes may be required to relocate to other properties or apartments due to necessary repairs and restoration. The return of the move-out rates to that of normal levels indicates that people stop switching homes. Accordingly, we used location-based big data with disaster impact and socio-demographic data to: (1) assess the duration of the evacuation and the time for the move-out rates to return to steady state after a hurricane, and (2) in responding to disasters, reveal potential disparities of evacuation return and home-switch stability on income, race, ethnicity. The remainder of this paper proceeds as follows: Section \ref{sec:Results} explains the results of the evacuation and home move-out rates and their corresponding time to return after the disaster, Section \ref{sec:Discussion} discusses and concludes the disparate recovery patterns of different sub-populations and the main contribution of this study, and Section \ref{sec:Methods} introduces the data and methods used in this study to assess evacuation return and home-switch stability.

\section{Results}
\label{sec:Results}
This section presents the analysis results of the evacuation and move-out rates addressing two research objectives: (1) to quantify and evaluate the duration of evacuation and the time for the move-out rates to return to steady state after a hurricane; (2) to evaluate and identify patterns experienced by low-income and racial/ethnic minority subpopulations.
\subsection{Duration of Evacuation and Move-Out Rates Returning to Steady State}
\begin{figure}
	\centering
    \includegraphics[width=0.85\linewidth]{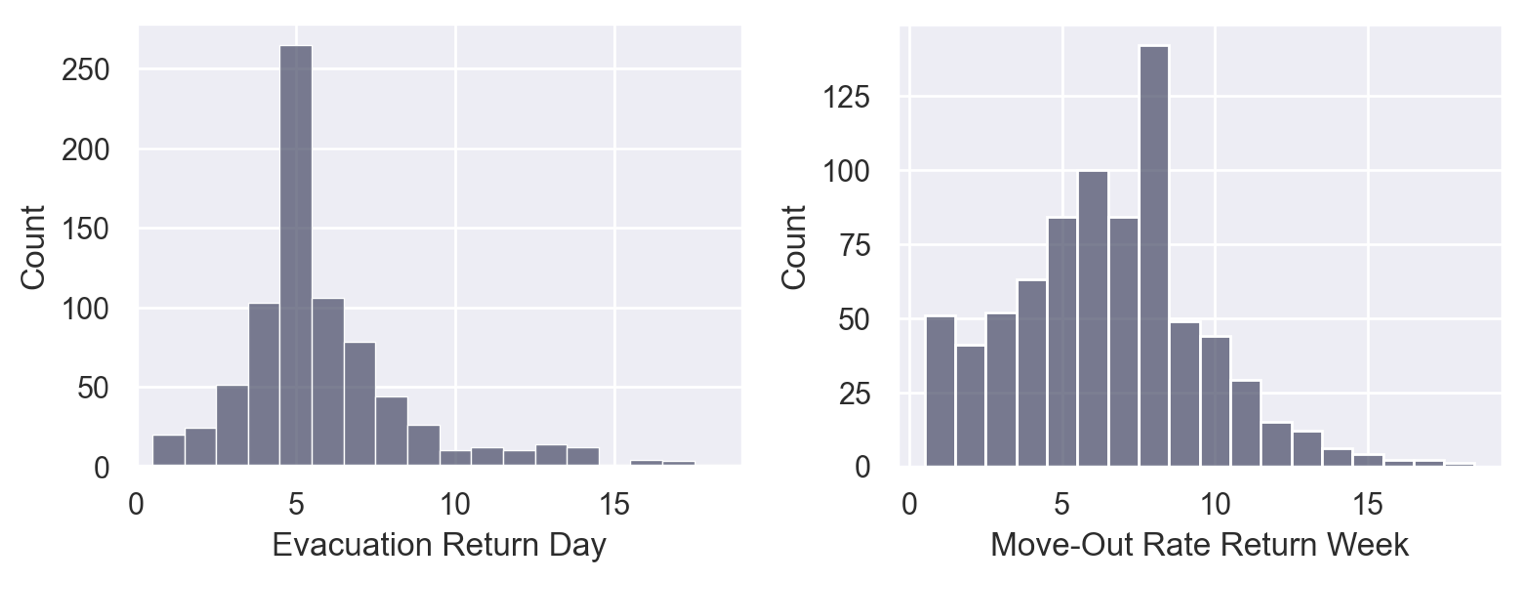}
    \caption{Distributions of the duration of the evacuation return (left) and the returned move-out rate (right) in Harris County, where the count represents the number of census tracts.}
	\label{fig:fig2}
\end{figure}

Figure \ref{fig:fig2} shows the distributions of the duration of the evacuation return and the returned move-out rate in Harris County, where the count represents the number of census tracts. According to the results of the evacuation rate, residents of more than half of the census tracts in Harris County stopped evacuating and were able to return to their homes within five days after landfall of Hurricane Harvey. The ability to return home may also indicate the extent of impacts, such as road inundations and power outages, had been reduced to acceptable levels. On the other hand, for the return duration of the move-out rate, the results show that people living in more than half of the census tracts in Harris County stopped moving out their homes within six weeks after landfall of Hurricane Harvey. The return duration for the move-out rate takes longer than the evacuation rate due to the nature of switching homes. Move-out rate is affected by more factors than evacuation. Insurance for covering costs of home restoration and whether to continue living in an area that is susceptible to flowing are a consideration for homeowners. Thus, the return duration of the move-out rate is more dispersed temporally than the evacuation rate since the reasons leading to relocate are more diverse. Despite that, the move-out rate of most of the census tracts returned to steady state within 8 weeks, and only a few census tracts have the return durations for the move-out rate longer than 9 weeks. Overall, most of the census tracts in Harris County took less than 5 days to return to steady state for evacuation and, got move-out rates, 8 weeks.

\subsection{Return Duration in Flooded and Non-flooded Areas}
\begin{figure}
	\centering
    \includegraphics[width=0.8\linewidth]{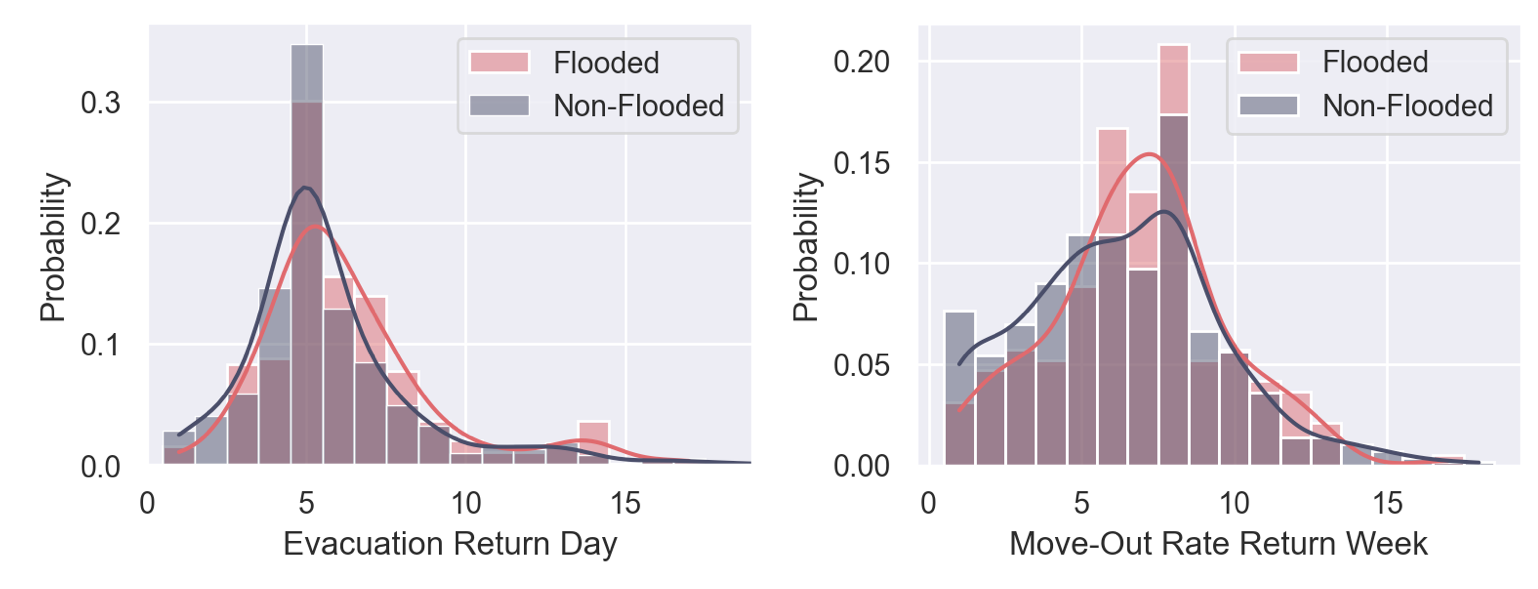}
    \caption{Distributions of the duration of the evacuation return (left) and the returned move-out rate (right) in flooded and non-flooded census tracts in Harris County, where the probability represents the percentage of census tracts. The p-values of the evacuation and the move-out rates between flooded and non-flooded census tracts are 0.0001 and 0.028, respectively. The results indicate that the durations for the flooded census tracts for both evacuation and move-out rates were significantly longer than for the non-flooded census tracts.}
	\label{fig:fig3}
\end{figure}

This section presents the results of the duration of returning to steady state with the consideration of flood impacts. Based on flood impact data, we categorized census tracts into two groups: flooded and non-flooded. Due to different degrees of impact, the patterns of the evacuation and move-out rates are likely to be different. For example, the evacuation return duration for people living in severely flooded areas may be of longer duration due to the wait for water to recede from their homes and remediation to be completed. Due to the non-normality of the residuals, this study used the Kruskal-Wallis test, or one-way ANOVA, on ranks, to examine the difference in the return patterns of different groups of populations. A significant result of the Kruskal-Wallis test indicates that the median values of different groups of census tracts are different. As shown in Figure \ref{fig:fig3}, the probability for relatively long return durations of evacuation and move-out rates in the flooded census tracts is generally higher than in the non-flooded census tracts. In addition, the p-values for both comparisons are less than 0.05, indicating that the return durations in the flooded census tracts for both evacuation and move-out rates were significantly longer than in the non-flooded census tracts.

\subsection{Examination of Disparities in Evacuation Return and Home-switch Patterns}
After we identified the durations of the evacuation return and the returned move-out rate, we analyzed whether the patterns of achieving short-term critical recovery milestones among various socio-demographic statuses were different. For example, the low-income population may experience difficulty evacuating from their damaged homes due to a lack of resources and the need for assistance from agencies or rescue organizations. Specifically, in the case of flooding, for people living in flooded or high-humidity houses without evacuating, the possibility of contracting viral diseases and infections could increase. The high-income population, however, may have the means to evacuate to temporary shelters, such as hotels and the homes of friends or relative, to mitigate physical impacts. The analysis results in this section examines socio-demographic status: median household income, the ratio of Black and Hispanic populations to total population, and the percentage of people living in rental homes, with the flooded impact data and the specified critical recovery milestones, evacuation return and home-switch stability, to understand the recovery patterns in different subpopulations and communities. Also, we investigated the differences between long and short return durations in terms of socio-demographic status.\\

\begin{figure}
	\centering
    \includegraphics[width=0.85\linewidth]{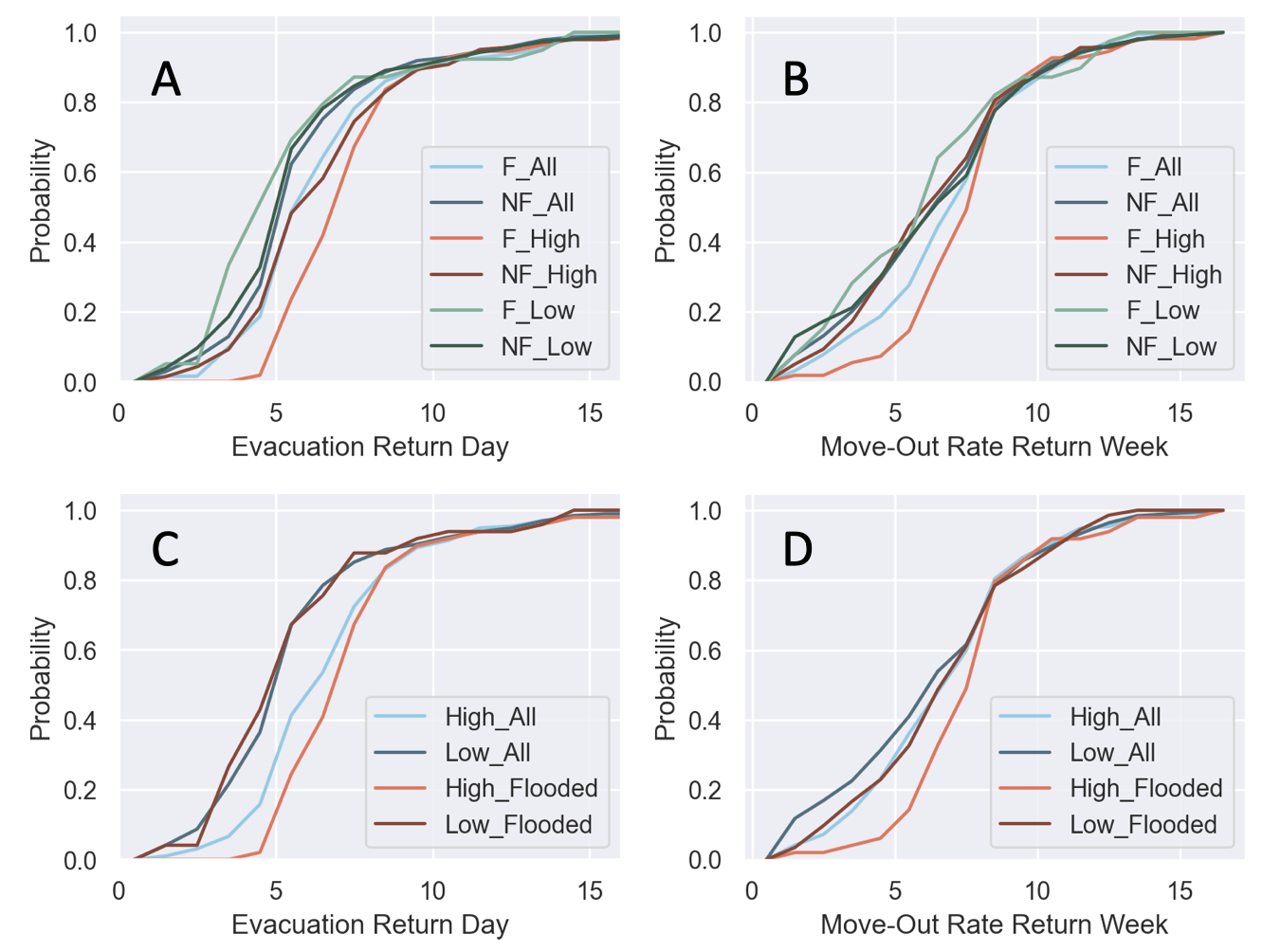}
    \caption{Return patterns of the evacuation and move-out rates for different subpopulations. A and B demonstrate the comparison between flooded (F) and non-flooded (NF) areas in all, high-, and low-income populations. C and D demonstrate the comparison between high- and low-income populations in all and flooded census tracts.}
	\label{fig:fig4}
\end{figure}

Persons of different socio-demographic status may exhibit different return patterns. The return patterns of census tracts affected by flooding may differ from those not impacted by flooding. Specifically, we compared the return patterns among combinations of high and low median-household-income levels, as well as flooded and non-flooded. Figure \ref{fig:fig4} illustrates the comparison results in two settings for the evacuation and move-out rates: (1) classify population into three groups, which are all populations, high income population (above the third quartile), and low income population (below the first quartile) and compare the patterns of flooded and non-flooded areas in each subpopulation group, as shown in Figures \ref{fig:fig4}A and \ref{fig:fig4}B, and (2) classify population into two groups, which are all population and population in flooded areas and compare the patterns of high and low median household income in each subpopulation group, as shown in Figures \ref{fig:fig4}C and \ref{fig:fig4}D.
In Figures \ref{fig:fig4}A and \ref{fig:fig4}B, the differences between flooded and non-flooded areas are significant in all populations and high-income populations. That is, the return durations of the evacuation and home move-out rates were longer in the flooded areas for all populations and high median household income areas. For the low-income subpopulation, however, the difference between flooded and non-flooded areas are not significant for both the evacuation and move-out rates. In other words, the flood impacts did not significantly affect the return patterns of the evacuation and move-out rates in the low-income population. In Figures \ref{fig:fig4}C and \ref{fig:fig4}D, the differences are significant between high- and low-income populations in the flooded areas for both the evacuation and move-out rates. For the comparison of high- and low-income populations in all study areas, only the evacuation rate shows significant differences between high- and low-income levels. The difference in the return patterns between the high- and low-income populations in all study areas for the move-out rate is insignificant. In addition, all return patterns show two-phase return progress that the return pace of the first 80\% of the population is faster than the remaining 20\% of the population, which indicates that most of the communities return to steady state earlier and 20\% of the areas return with considerable lag.\\

According to these analysis results, in flooded census tracts in the study area, the time for the evacuation and home move-out rates to return to steady state was longer. The longer return duration compared to the non-flooded census tracts is intuitive: residents in the flooded areas must wait for the flooding to recede before returning to homes from evacuation and making relocation decisions. Yet the exception is for the low-income subpopulations; the differences between flooded and non-flooded status for the low-income census tracts are insignificant for both the evacuation and move-out rates. This may show the inability of low-income population to evacuate and relocate to mitigate the impact of the flooding. For the flooded areas, it is significant that the low-income population has a shorter return duration for both the evacuation and move-out rates than the high-income population. A further comparison between long and short return durations of the evacuation and move-out rates is addressed in the following section. Overall, based on the analysis of the two indicators of short-term critical recovery milestones, the immediate responses and the community recovery progresses in terms of evacuation and home-switch are different between high- and low-income populations, as well as between flooded and non-flooded areas. 

\subsection{Comparisons Between Long and Short Return Duration}
\begin{figure}
	\centering
    \includegraphics[width=0.95\linewidth]{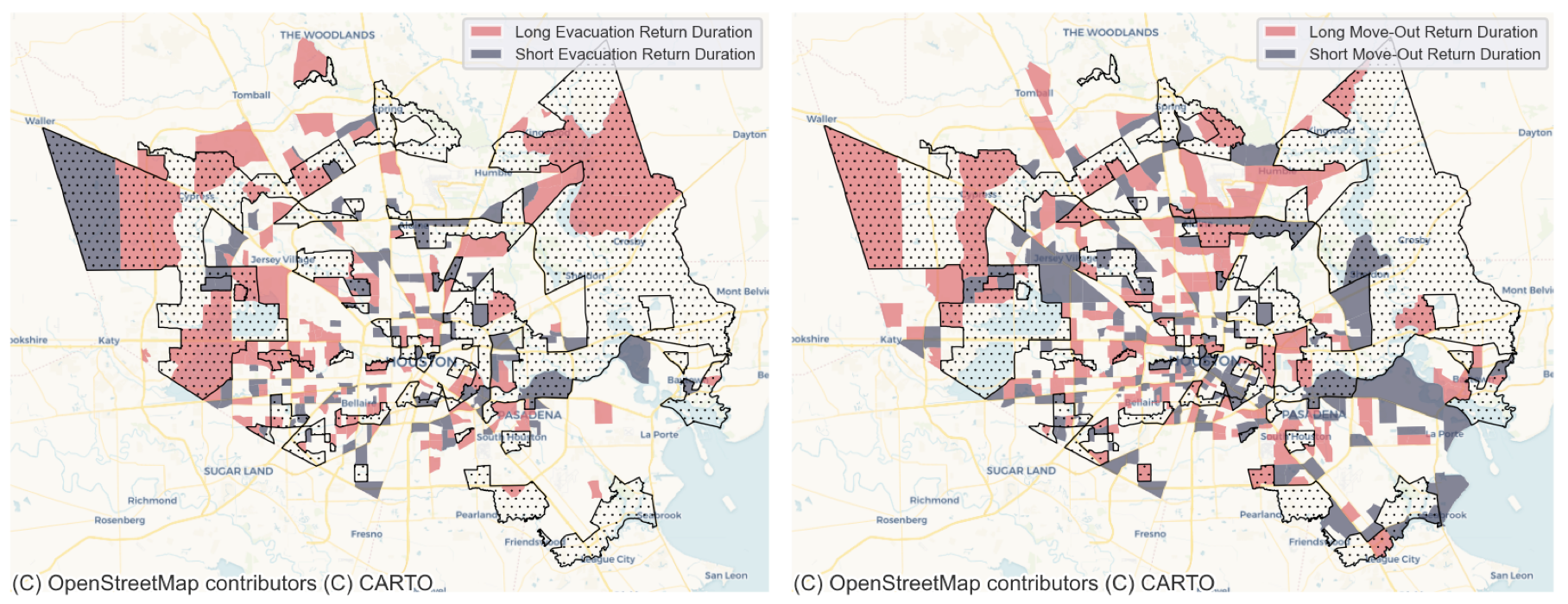}
    \caption{Locations of census tracts with long and short return durations in terms of the evacuation rate (left) and the home move-out rate (right). The dotted areas are the census tracts identified as flooded census tracts.}
	\label{fig:fig5}
\end{figure}

The results in the previous section examined the effect of income-level and flooding status on the return patterns on the achievement of short-term critical recovery milestones of the evacuation and move-out rates. To understand the difference between long and short return durations related to the evacuation and move-out rates, we compared these milestones with respect to income, housing type, and race. Based on the distribution of return durations for the evacuation and move-out rates, the first quartile durations of return to steady state in flooded areas is 5 days for evacuation and 5 weeks for move-out rates. The third quartile values are 7 days for evacuation and 8 weeks for move-out rates. We applied the following criteria to distinguish long and short return durations. Long return duration was greater than 7 days for evacuation and 8 weeks for home move-out rates. Short return duration was less than 5 days for evacuation and less than 5 weeks for home move-out rates. Figure \ref{fig:fig5} shows the location of census tracts with long and short return durations in terms of the evacuation and home move-out rates. 

\begin{figure}
	\centering
    \includegraphics[width=0.85\linewidth]{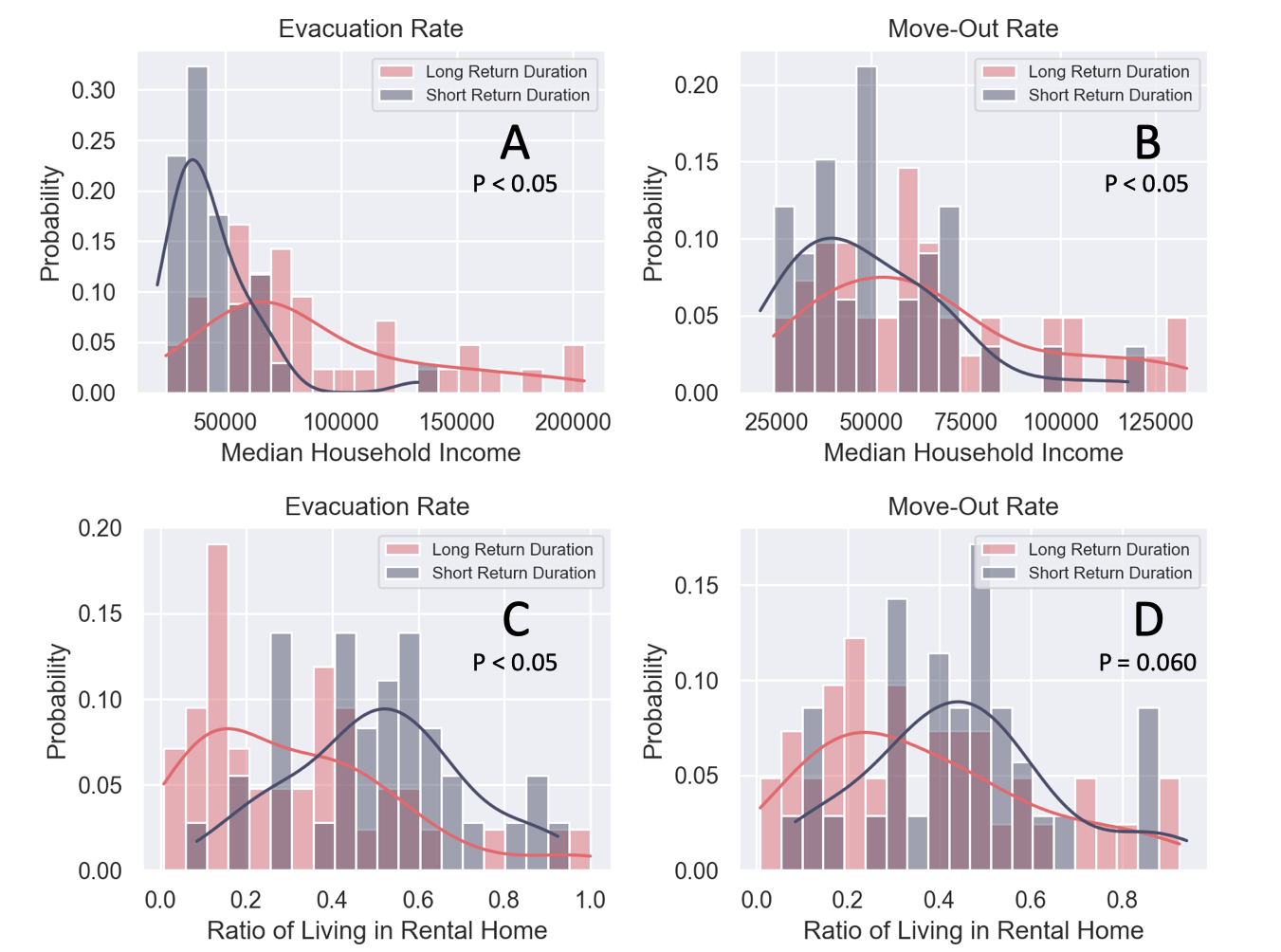}
    \caption{Comparison between long and short return durations with respect to the median household income (A and B) and the ratio of the population living in rental homes (C and D) of the evacuation (A and C) and move-out rates (B and D). The statistical test results indicate that the differences between long and short return durations are significant except for the difference in the move-out rate in terms of the ratio of living in renting homes.}
	\label{fig:fig6}
\end{figure}

We first compared the differences between long and short return durations with respect to median household income and the ratio of the population living in rental home. Figure \ref{fig:fig6} compares long and short return durations to the median household income and the ratio of the population living in rental homes in the evacuation and move-out rates in the flooded census tracts. The statistical test results indicate that the difference between long and short return durations are significant except for the difference in the move-out rate in terms of the ratio of population living in rental homes; even though it is insignificant, it is apparent that the census tracts with long move-out return duration tend to have a lower ratio of residents living in rental properties. Thus, according to this result, the census tracts with lower median household income and a higher ratio of persons living in rental homes had short return durations in the evacuation and move-out rates.
The shorter evacuation rate return duration for the low-income population may indicate the inability to evacuate, which is reported in other studies in the literature \citep{renne_carless_2011}. Several studies \citep{elder_african_2007, thompson_evacuation_2017, toledo_analysis_2018} indicated that low-income population were less likely to evacuate due to barriers linked with financial constraints. For example, the high-income population can evacuate to other cities or hotels. In contrast, the low-income population may not have options other than staying in their homes. The shorter home move-out rate return duration for the low-income population compared to the high-income population may be related to their housing types (renting versus owning), the damage level of their homes, and the capability to repair their homes (influenced by flood insurance coverage). The census tracts with low median household income usually have a higher ratio of the population of living in rental apartments/homes. The residents may be required to move to other properties or apartment units because of flooding damages. The high-income population is able to live in houses less vulnerable to flooding \citep{deria_evaluating_2020} or have flood insurance and financial resources \citep{atreya_what_2015, browne_demand_2000}, allowing them more time to consider and make decisions regarding rebuilding their home and relocation. People unable to afford repair costs or living in rental apartments/homes are less likely to return to their homes due to insufficient financial resources \citep{fussell_homeownership_2014}. Thus, a shorter move-out return duration does not necessarily signal a positive trend in short-term recovery and these indicators should be interpreted in light of the housing type and socio-demographic characteristics of each area.\\

\begin{figure}
	\centering
    \includegraphics[width=0.85\linewidth]{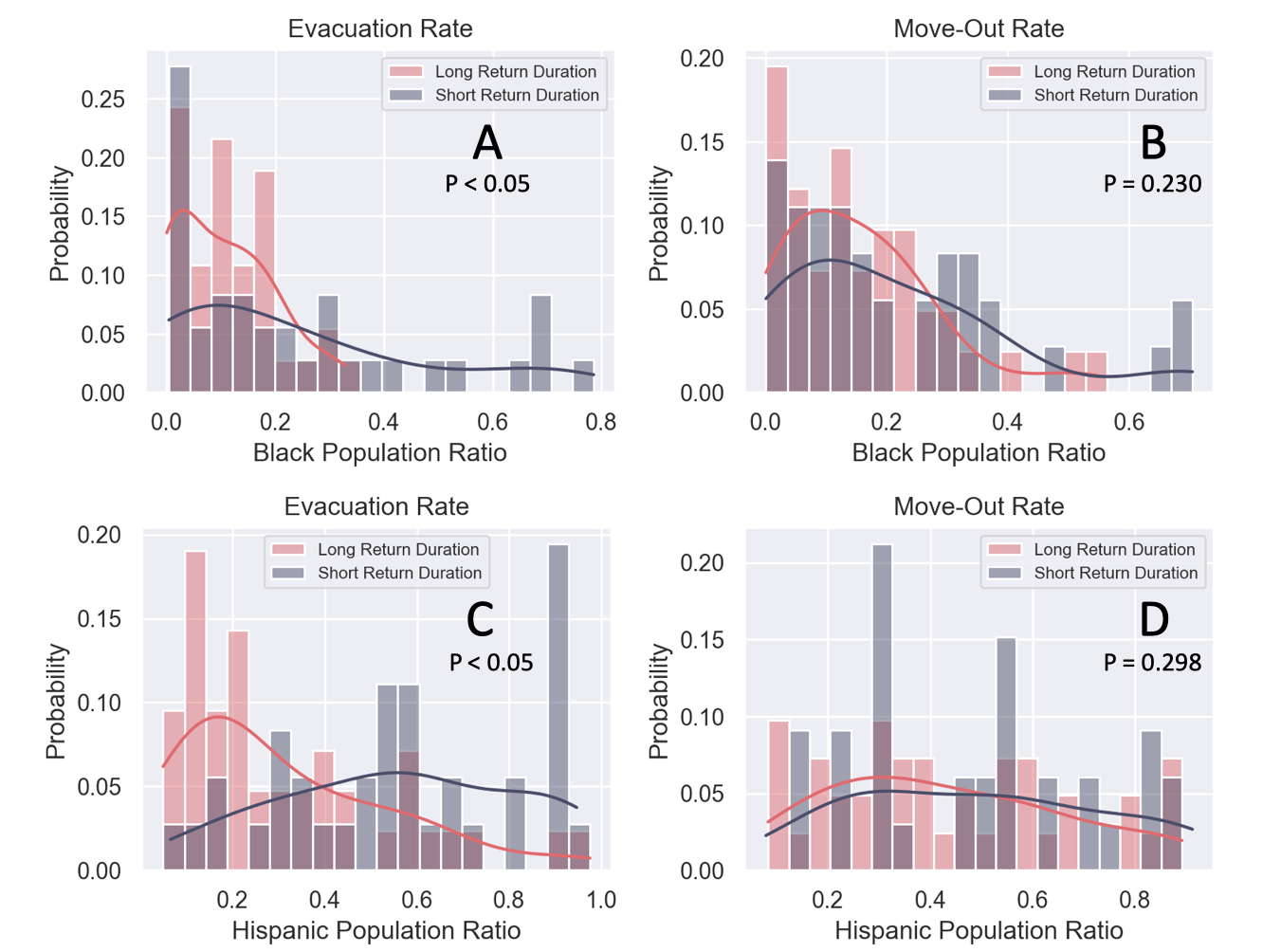}
    \caption{Comparison between long and short return durations with respect to the Black population ratio (A and B) and Hispanic population ratio (C and D) in the evacuation (A and C) and move-out rates (B and D). The statistical test results indicate that the differences between long and short return durations in the evacuation rate are significant but insignificant in the home move-out rate.}
	\label{fig:fig7}
\end{figure}

Figure \ref{fig:fig7} compares long and short return durations in terms of the Black and Hispanic populations in the evacuation and move-out rates in the flooded census tracts. The statistical test results indicate that the differences between long and short return durations in the evacuation rate are significant but for the home move-out rate, are insignificant. Since the evacuation rate indicates the immediate response of the population to the disaster, an ideal result of the difference between long and short return duration should be solely affected by the flood damage level and not related to the socio-demographic. However, a census tract with a faster evacuation return duration tends to have a higher ratio of minority populations. This result indicates that minority populations require help and resources to overcome challenges to evacuation. On the other hand, based on the statistical results, there is no significant difference in the move-out return duration with respect to the ratio of minority populations.

\section{Discussion and Concluding Remarks}
\label{sec:Discussion}
By employing location-based big data with disaster impact and socio-demographic data, this study specified two critical recovery milestones, evacuation return and home-switch stability, to address the objectives of the paper. Specifically, we used the evacuation rate to assess results demonstrating that more than half of the census tracts in Harris County returned from evacuation within 5 days. In addition, the populations more than half of the census tracts stopped moving out after 6 weeks. Return durations of the high-income census tracts when flooded were longer than those for when non-flooded; however, there was no significant difference between flooded and non-flooded in the low-income census tracts. This finding indicates the inability of the low-income population to evacuate and relocate. When the census tracts were flooded during Hurricane Harvey, the disparate return patterns in the evacuation and home move-out rates were significant in that the low-income population returned sooner than the high-income population. The flooded census tracts with short evacuation return (less than 5 days) had lower median household income, higher ratio of persons living in rental homes, and a higher percentage of minority populations compared to those with long evacuation returns. On the other hand, there is no significant difference between long and short return durations of the home move-out rate in the flooded census tract with minority groups; the differences between them are mainly related to income and housing types. The long return durations (more than 8 weeks) of the home move-out rate tended to be more prevalent in high-income census tracts compared to the short return durations. Often, we view the areas with shorter return durations as more resilient to disaster based on the ability to return to steady state. According to the results and discussions in this study, however, the fact might be just the opposite. A shorter evacuation return and relocation progress may indicate that challenges faced by low-income and minority populations to evacuate and relocate and require additional assistance and resources to mitigate the impact of disasters. \\

This study provides three contributions to the study of parity recovery and remediation after a disaster: specifying critical short-term recovery milestones, revealing disparate community recovery patterns in different subpopulations, and observing non-uniform recovery duration and patterns. First, this study specified two critical milestones—evacuation return and home-switch stability—related to short-term disaster recovery to be used for more data-driven and proactive monitoring of recovery. The standard approaches for community recovery monitoring have significant lags and put the burden of data collection on affected people via public surveys. The indicators used in this study were obtained from privacy-protective and aggregated location-based data to provide a finer-resolution insight into the recovery at the census-tract level and to monitor community recovery progress in a more data-centric manner during and in the aftermath to support decision-makers and responders proactively. Second, the findings showed that a shorter duration of critical recovery milestone indicators in flooded areas is not necessarily a positive indication. A short duration of evacuation return could be due to challenges to evacuation faced by low-income residents. A short home move-out return could be due to living in rental property or a lack of flood insurance to properly effect home repairs and relocation. In practice, early return of evacuation and home-switch in the context of flooding events may signal the absence of resources and may require support from officials and decision-makers. Third, the skewed distribution of return durations for both the evacuation return and home-switch stability were observed in all subpopulation groups. All return patterns show a two-phase return process that the first 80\% of population returned faster than the remaining 20\% of the population. This phenomenon indicates that the recovery patterns are non-uniform. Hence, in evaluation and monitoring of recovery, it is important to consider the socio-demographic information of both unusual short and long return duration to identify potential issues before making decisions. From a practical perspective, the indicators and findings in this study could inform disaster managers and public officials to make recovery decisions and allocate resources in a more proactive, data-driven, and equitable manner. Such data-driven approach could overcome lags and inefficiencies in disaster recovery management and enhance community resilience.

\section{Materials and Methods}
\label{sec:Methods}
\subsection{Study Area and Period}
The study collected and analyzed data from Harris County, Texas, which includes the Houston metropolitan area, one of the most adversely affected areas by the 2017 Hurricane Harvey. On August 25, 2017, Hurricane Harvey, a devastating Category 4 hurricane, made landfall and led to heavy rainfall in Harris County. In addition, Houston downtown and some western areas of Harris County were flooded. There was limited mandatory evacuation issued for Harris County. Also, due to the release of water from Barker and Addicks Reservoir, the west part of Harris County experienced an extensive and prolonged flooding. The impacts of Hurricane Harvey continued until September 1, 2017, when Hurricane Harvey left Harris County, and people started to recover from the impacts afterward. To understand the progress of return on evacuation and home switch, we obtained data from the period between July 9, 2017, to July 28, 2018, within which the pre-disaster period is from July 9 to August 24, 2017, and the post-disaster period is from August 25 to July 28, 2018. 
\subsection{Data Sources}
\subsubsection{Location-based Data}
The aggregated location-based data used in this study is from Cuebiq Inc. Cuebiq has a location intelligence platform which collects mobility data of anonymized devices of users who have opted in to provide access to their location data anonymously for research purposes through a CCPA- and GDPR-compliant framework. Cuebiq creates their geo-behavioral dataset by collaborating with app developers directly to capture offline behavior data at fine-granular scales with accurate locations based on Bluetooth technology, as well as GPS, Wi-Fi, and IoT signals. For anonymous, opted-in each user, Cuebiq collects more than a hundred data points daily on average, which provides a more accurate understanding of the population’s mobility patterns than the conventional mobility survey. Current daily active user count collected by Cuebiq is roughly 15 million in the United States. Through its Data for Good program, Cuebiq provides mobility insights for academic research and humanitarian initiatives. By analyzing the aggregated mobility patterns of more than 500,000 anonymous Cuebiq users (representing 12.5\% of the population of the Puget Sound region under analysis), \citet{wang_extracting_2019} determined that Cuebiq data, as compared to cellular network and in-vehicle GPS data, benefitted from a superior combination of large scale, high accuracy, precision, and observational frequency. Beyond validating scale and accuracy, the research \citep{wang_extracting_2019} found that Cuebiq data is highly demographically representative. In addition, multiple existing studies \citep{aleta_modelling_2020, nande_effect_2021} on Cuebiq data have demonstrated the representativeness of the data.\\

Cuebiq aggregates data using artificial intelligence and machine learning techniques. Cuebiq's responsible data sharing framework enables us to query anonymized, aggregated, and privacy-enhanced data, by providing access to an auditable and on-premise sandbox environment. In this study, we used one of the Cuebiq aggregated datasets, Daily Metric by Device, to assess the return patterns and progresses after disasters. The Daily Metric by Device table provides information at the device level, including users’ home census block groups and hours users stay at home census block groups per day and night. Based on the aggregated data, we calculated the evacuation and home move-out rate, which are introduced in Data Processing section, to understand population’s response to disasters. All the location-based data used in this study were aggregated to the census tract level in order to further preserve privacy.

\subsubsection{Flood Impacts Data}
In this study, we used flood inundation percentages within a census tract as a measure of flood impacts. Specifically, we calculated flood inundation percentages based on the flood inundation map of Hurricane Harvey produced by Federal Emergency Management Administration (FEMA). We overlaid the map of Harris County at the census tract level with the FEMA flood inundation map to compute the flood inundation areas within each census tract and its corresponding flood inundation percentage. For the analysis in this study, we used 10\% of the flood inundation percentage in each census tract as a threshold to distinguish flooded and non-flooded areas. Figure 8 shows the spatial distribution of the flooded and non-flooded census tracts in Harris County. 
\begin{figure}
	\centering
    \includegraphics[width=0.6\linewidth]{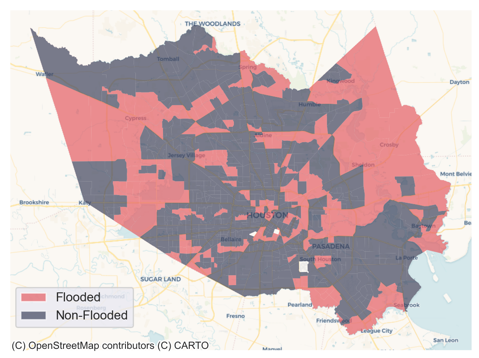}
    \caption{Map of the flooded and non-flooded census tracts in Harris County based on the flood inundation map of Hurricane Harvey produced by FEMA.}
	\label{fig:fig8}
\end{figure}
\subsubsection{Socio-demographic Data}
We retrieved demographic and household socioeconomic data from the American Community Survey database administrated by US Census Bureau at the census tract level to understand whether the socio-demographic status affects return patterns and progresses. The data used in this study is the 2017 5-year estimates data, representing the estimates over the five-year period from 2013 through 2017. The socio-demographic data obtained in this study included the median household income, the ratio of the Black population, and the ratio of the Hispanic population in each census tract in Harris County. We then compared this socio-demographic data with the recovery milestones such as evacuation and home switch return patterns.
\subsection{Data Processing}
We first identified residents of Harris County during the hurricane period based on their home census block group tags in the aggregated data provided by Cuebiq. That is, we extracted all users and their corresponding information if their home tags occurred in Harris County at least once. For the evacuation rate, we calculated the percentage of people who left their homes daily. Also, we elicited the move-out rate from the changes of users’ home tags in the aftermath of the event. Calculation of the evacuation and move-out rates is discussed in the following sub-sections.
\subsubsection{Evacuation Rate}
The evacuation rate was calculated based on the data at the census block group level, the finest geospatial level of users’ home locations. In this study, people who left their home census block groups and dwelled in the other census block groups for an entire day were viewed as evacuated populations. In other words, the evacuation rate indicates that the percentage of people who left their home census block groups during Hurricane Harvey. To this end, we extracted the data from July 9, 2017, through November 19, 2017, from the aggregated data provided by Cuebiq to capture the pre-disaster period and avoid the effect of the Thanksgiving holiday. To ensure the quality of the data and avoid data biases, we analyzed only records with at least 240 minutes of location information in a day. Then we calculated the evacuation rate for each census tract as the number of evacuated users divided by the total number of users in a census tract. Also, to understand fluctuations in the evacuation rate during Hurricane Harvey, we calculated the percent change of the evacuation rate according to the baseline rate for CBGs, which is the average evacuation rate of the pre-disaster period (July 9, 2017, through August 5, 2017). The baseline evacuation rate is calculated based on the day of a week to account for the difference between weekdays and weekends. The calculation of the percent change is shown as Equation (\ref{eq:eq1}).\\
\begin{equation}
	Percent \; Change \; of \; ER_{t,d,c}= {\frac {(ER_{t,d,c}-BER_{d,c})}{BER_{d,c}}}
	\label{eq:eq1}
\end{equation}
where, $ER_{t,d,c}$ is the evacuation rate on day $t$ and day $d$ of a week in census tract $c$, and $BER_{d,c}$ is the baseline evacuation rate on the day $d$ of a week in census tract $c$.

\subsubsection{Home Move-out Rate}
The home move-out rate, an index assessing home-switch return patterns and progress, is calculated according to the home CBG information in the aggregated data. Since home switches do not happen often, we obtained data from July 9, 2017, to July 28, 2018, and further aggregated it to a weekly period. Specifically, we aggregated the home tags of all users who had at least one home tag in Harris County to a weekly table so that every user has a home tag every week. If no identified home information can be found from Cuebiq for a user for a specific week, we filled the data from the previous home tag by assuming no home switch for this user during this period. We then used the home information and calculated the home move-out rate at the census-tract level. To this end, the move-out rate is defined as the number of users switching their homes during a specific week from a census tract over the number of users in the census tract. Likewise, we calculated percent change of the move-out rate to understand fluctuations in the aftermath of Hurricane Harvey based on the pre-disaster baseline data from July 9, 2017, through August 12, 2017. Each census tract has a baseline home move-out rate for calculating percent changes using Equation (\ref{eq:eq2}).\\
\begin{equation}
	Percent \; Change \; of \; MOR_{w,c}= {\frac {(MOR_{w,c}-BMOR_c)}{BMOR_c}}
	\label{eq:eq2}
\end{equation}
where, $MOR_{w,c}$ is the move-out rate on week $w$ in census tract $c$, and $BMOR_c$ is the baseline move-out rate in census tract $c$.
\subsection{Identifying Return Duration}
The times at which the evacuation and home move-out rates return to steady state are critical milestones of community short-term recovery. Thus, we developed a duration identification approach to specify the duration of evacuation and move-out rates to return to steady state after the event. During the impact of a disaster, the evacuation rate increases as persons evacuate from their homes to avoid injury and due to physical damage and return to their homes when the impact level decreases. Similarly, people switch their homes during and in the aftermath of a disaster due to damaged property. The move-out rate increases directly following a disaster then returns to steady state. We used the percent change of the evacuation and move-out rates and applied the rolling average method to obtain trends of the percent change of the evacuation and move-out rates during the evaluation periods. In particular, we calculated the seven-day average changes for the evacuation rate when there were more than 4 days of data within each 7-day period. Also, to understand fluctuations in the move-out rate, we used 4-week average changes in cases in which at least two weeks of data were known. \\

After determining trends in the percent change of the evacuation and move-out rates for each census tract in the study area, we looked for the maximum change and the durations of reaching the maximum change during the evaluation period. Using that information, we identified the duration for return to steady state in each census tract in the period after the maximum change and then specified the beginning of a steady state. Since a new steady state in the aftermath of disasters can vary compared to the pre-disaster status, we defined a steady state as having no substantial difference in terms of the percent changes. That is, the difference of the percent change of the evacuation and home move-out rates between two consecutive days/weeks is within a threshold, assumed to be 10\% in this study. Thus, we defined the returned day/week of the evacuation and move-out rates for a census tract to be the first day/week with two consecutive differences less than 10\%. Through this approach, we identified the duration of the evacuation and move-out rates returning to steady states for each census tract in Harris County in the aftermath of Hurricane Harvey.

\section{Data availability}
The data that supports the findings of this study are available from Cuebiq, but restrictions apply to the availability of the data, which were used under license for the current study. The data can be accessed upon request submitted on Cuebiq. The socio-demographic and flood impact data used in this study are all publicly available through the data repositories of the U.S. Census Bureau and Federal Emergency Management Agency.
\section{Code availability}
The code that supports the findings of this study is available from the corresponding author upon request.
\section{Acknowledgements}
This material is based in part upon work supported by the National Science Foundation under Grant CMMI-1846069 (CAREER), and the Texas A\&M University X-Grant 699. The authors also would like to acknowledge the data support from Cuebiq. Any opinions, findings, conclusions or recommendations expressed in this material are those of the authors and do not necessarily reflect the views of the National Science Foundation, Texas A\&M University, or Cuebiq.

\bibliography{ref}  






\end{document}